\documentclass{fkfs}

\IfFileExists{latin1.sty}{\usepackage{latin1}}{\usepackage{isolatin1}}

\usepackage[pdftex]{graphicx}

\usepackage{amsmath}

\usepackage[caption=false,font=normalsize,labelfont=sf,textfont=sf]{subfig}

\usepackage{stfloats}
\usepackage{booktabs}
\usepackage{multirow}
\usepackage{array}
\usepackage{longtable}
\usepackage{tabularx}
\usepackage{spreadtab}
\usepackage{siunitx}
\usepackage{tabu}
\usepackage{ltxtable}
\usepackage{filecontents}
\usepackage{colortbl}
\usepackage{hhline}
\usepackage[rgb,dvipsnames,table]{xcolor}
\usepackage{amssymb}
\usepackage{rotating}
\usepackage{adjustbox}
\usepackage{csquotes}
\usepackage{lipsum}
\usepackage{tikz}

\usepackage{hyperref}
\hypersetup{
	bookmarks=true,         % show bookmarks bar?
	unicode=false,          % non-Latin characters in Acrobat?s bookmarks
	pdftoolbar=true,        % show Acrobat?s toolbar?
	pdfmenubar=true,        % show Acrobat?s menu?
	pdffitwindow=false,     % window fit to page when opened
	pdftitle={Publication},    % title
	pdfauthor={Florian Hauer},     % author
	pdfsubject={Software Engineering},   % subject of the document
	pdfcreator={Florian Hauer},   % creator of the document
	pdfproducer={},  % producer of the document
	pdfkeywords={System Verification} {Autonomous Driving} {Search-Based Testing}, % list of keywords
	pdfnewwindow=true,      % links in new window
	colorlinks=false,       % false: boxed links; true: colored links
	linkcolor=red,          % color of internal links
	citecolor=green,        % color of links to bibliography
	filecolor=magenta,      % color of file links
	urlcolor=cyan           % color of external links
}

\newcolumntype{L}[1]{>{\hsize=#1\hsize\raggedright\arraybackslash}X}%
\newcolumntype{R}[1]{>{\hsize=#1\hsize\raggedleft\arraybackslash}X}%
\newcolumntype{C}[1]{>{\hsize=#1\hsize\centering\arraybackslash}X}%

\renewenvironment{quote}%
{\list{}{\leftmargin=0.2in\rightmargin=0.2in}\item[]}%
{\endlist}

\author{Florian Hauer \\ Technische Universität München \\ florian.hauer@tum.de
	\and
	Bernd Holzmüller \\ ITK Engineering GmbH \\ bernd.holzmueller@itk-engineering.de
}
\title{Szenario-Optimierung für die Absicherung \\von automatisierten und autonomen Fahrsystemen}
\begin{document}
\maketitle

\begin{abstract}
Die Verifikation und Validierung automatisierter und autonomer Fahrsysteme, besonders das Finden geeigneter Testszenarien für die virtuelle Absicherung, stellen eine enorme Herausforderung dar.

Diese Arbeit stellt eine Testmethodik vor, die metaheuristische Suche adaptiert, um Szenarien zu optimieren. Hierfür muss ein passender Suchraum und eine geeignete Gütefunktion aufgestellt werden. Ausgehend von einer abstrakten Beschreibung der Funktionalität und Anwendungsfälle des Systems werden parametrisierte Szenarien abgeleitet. Deren Parameter spannen einen Suchraum auf, aus dem die passenden Szenarien zu identifizieren sind. Hierfür werden suchebasierte Verfahren verwendet, die sich an einer Gütefunktion orientieren und somit die Szenarien identifizieren, in denen das System das schlechteste Verhalten zeigt. Bei geeigneter Ableitung einer solchen Gütefunktion kann die Basis geschaffen werden für eine Argumentation über Vollständigkeit der Tests und über die Qualität des Systemverhaltens. Außerdem wird ein zielgerichtetes Testen mit automatischer Testfallauswertung ermöglicht.% worauf zu achten
\end{abstract}

\section{Motivation}
Das Streben nach autonomen Fahrsystemen resultiert in immer komplexeren und umfangreicheren Systemen. Prototypen unterschiedlicher Automatisierungsgrade und verschiedener Unternehmen sind in Form von Testfahrzeugen im Realverkehr unterwegs und erzielen vielversprechende Ergebnisse.

Dennoch gibt es auf dem Weg zur Serienreife noch einige Hürden. Besonders die Absicherung solcher Systeme stellt aufgrund der Systemkomplexität und der potentiell unendlichen Szenarien eine große Herausforderung dar \cite{koopman2017autonomous}. Auch wenn Realtestfahrten weiterhin sehr wichtig für Verifikation und Validierung sein werden, reicht die reale Absicherung allein nicht aus \cite{wachenfeld2016release}\cite{helmer2015safety}\cite{zhaoandpeng}. Stattdessen soll verstärkt virtuell in einem szenariobasierten closed-loop Aufbau getestet werden \cite{ulbrich2017testing}. Die Identifikation und Wahl von passenden Testszenarien ist extrem schwierig und eine der wesentlichen Problemstellungen. 

Da keine konkrete und präzise Beschreibung existiert, was ein autonomes Fahrzeug im Detail können muss \cite{li2016intelligence}, kommt es zu vielen Herausforderungen bei der Testfallerzeugung, -auswahl und -auswertung \cite{briand2016testing}\cite{stellet2015testing}. Aktuelle Testmethodiken reichen nicht aus \cite{bengler2014three}\cite{helle2016testing} und Verbesserungen sind notwendig \cite{hendriks2010future}\cite{huang2016autonomous} (siehe Abschnitt \ref{sec:currentapproaches}). Spätestens bei höherautomatisierten Fahrsystemen werden Teile der Funktionalität von maschinell erlernten Funktionen bereitgestellt, welche mit inhärenten Schwierigkeiten bei der Whitebox-Verfikation verbunden sind \cite{burton2017making}. Somit liegt der Fokus vor allem auf Blackbox-Verfahren, wofür suchebasierte Verfahren häufig als vielversprechende Technologie sowohl in Theorie und Praxis vorgeschlagen werden \cite{sroka2014genetic}. Existierende Arbeiten zur Verwendung von suchebasierten Verfahren für den Test von Fahrfunktionen fokussieren sich auf die technologischen Aspekte und nehmen den für die Suche vorausgesetzten Suchraum und die passende Gütefunktion als gegeben an oder erstellen diese ad-hoc (siehe Abschnitt \ref{sec:currentapproaches}). Das lässt methodische Aspekte unberücksichtigt. Tatsächlich sind die Ableitung des Suchraums und der Qualitätsfunktion aber sehr schwierig, was bei nicht korrekter Durchführung in falschen Schlussfolgerungen über die Testergebnisse und die Sicherheit des Systems resultieren kann. Es ist außerordentlich wichtig, dass vielversprechende Technologien wie suchebasierte Verfahren in einer strukturierten Methodik angewandt werden. 

In dieser Arbeit wird eine solche Methodik für den Test von automatisierten und autonomen Fahrsysteme präsentiert. Dabei werden suchebasierte Verfahren adaptiert, indem die Mittel zur Ableitung von Suchräumen und Zielfunktionen gegeben werden. Ausgehend von abstrakten, funktionalen Beschreibungen wird beides, also Suchraum und Zielfunktion, erstellt. Sie dienen als Startpunkt für die heuristische Suche, um \enquote{gute} Testfälle zu erzeugen. So werden konstruktiv funktionalitätsspezifische Extremszenarien identifiziert, in denen das System das schlechteste Verhalten zeigt.

Abschnitt \ref{sec:scenariobasedtesting} beschreibt den Grundgedanken des szenariobasierten Testens. In Abschnitt \ref{sec:currentapproaches} wird ein Überblick über existierende Arbeiten gegeben, bevor anschließend in Abschnitt \ref{sec:sbapproaches} die heuristische Suche beschrieben wird. Nachdem in Abschnitt \ref{sec:methodology} die Methodik vorgestellt wurde, wird deren Anwendung für die Systemfreigabe in Abschnitt \ref{sec:release} diskutiert. In Abschnitt \ref{sec:conclusion} wird die Arbeit zusammengefasst.

\section{Szenariobasiertes Testen}\label{sec:scenariobasedtesting}

\subsection{Szenarien}
\begin{figure}[htb]
	\centering
	\begin{tikzpicture}
	\node (n1) at (0,0) {\includegraphics[width=10cm]{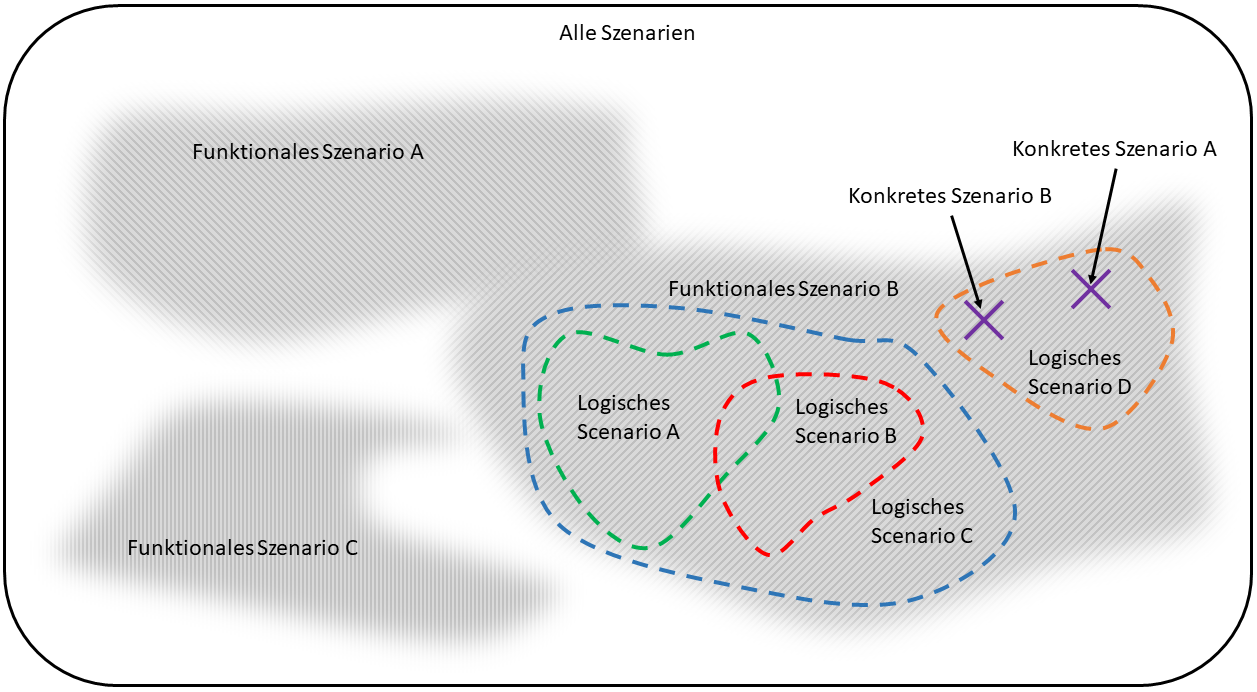}};
	\node (n2) at (8,-1.75) {\includegraphics[width=3cm]{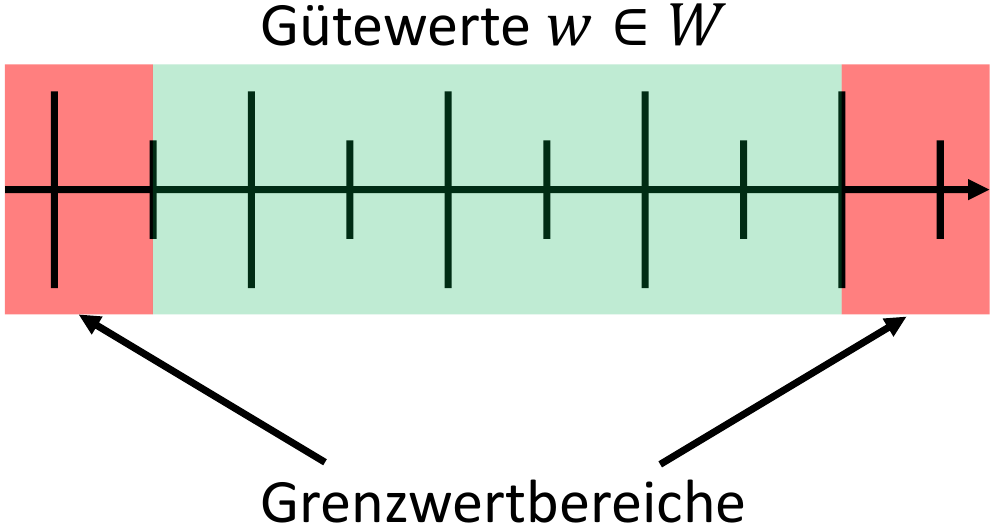}};
	\node (n3) at (3.55,0.4) {};
	\node (n4) at (9,-1.25) {};
	\draw[bend left,->]  (n3) to node [auto] {$q$} (n4);
	\node (n3) at (2.7,0.15) {};
	\node (n4) at (6.75,-1.25) {};
	\draw[bend left,->]  (n3) to node [auto] {$q$} (n4);
	\end{tikzpicture}
	\caption{Relationen zwischen funktionalen, logischen und konkreten Szenarien sowie die Zuordnung einer Güte zu einem konkreten Szenario}
	\label{fig:scenariosandquality}
\end{figure}
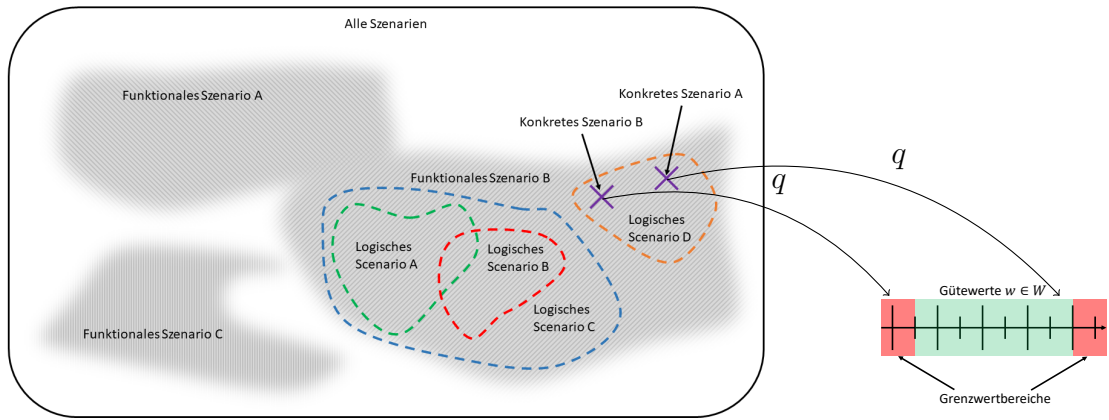

Das bereits verbreitete Konzept der funktionalen, logischen und konkreten Szenarien \cite{bagschikszenarien} wird für diese Arbeit aufgegriffen (siehe auch Abbildung \ref*{fig:scenariosandquality}). Funktionale Szenarien sind textuelle Beschreibungen von abstrakten Anwendungsfällen, in denen das System seine Funktionalität bereitzustellen hat \cite{bagschikszenarien}. Deren Detaillierung sind die logischen Szenarien, welche Betriebsszenarien durch Entitäten und deren Beziehungen mithilfe von Parameterbereichen darstellen \cite{bagschikszenarien}. Dies wird in dieser Arbeit wie folgt formalisiert: Ein logisches Szenario wird als Tripel $(D,V,I)$ definiert. Dabei enthält $D$ fixierte Informationen, welche den Rahmen der Szenarien darstellen. Die Variablen $v_i$ in $V$ ($\left|V\right|=n$) stellen Parameter dar, von denen jeder im zugehörigen Intervall $I_i \in I$ variiert werden kann. Wird jedem $v_i, i=1,..,n$ ein Wert zugeordnet, so erhält man eine Belegung für dieses logische Szenario. Eine solche Belegung entspricht einem konkreten Szenario. Diese sind die finalen, ausführbaren Szenarien \cite{bagschikszenarien}. Mithilfe der Intervalle spannen die logischen Szenarien einen potentiell unendlichen Raum $A=I_1 \times I_2 \times ... \times I_n \subset \mathbb{R}^n$ an möglichen konkreten Szenarien auf.

%\begin{figure}[htb]
%	\centering
%	\includegraphics[height=6cm]{figures2/scenarios.png}
%	\caption{TODO}
%	\label{fig:scenariosandquality}
%\end{figure}

\subsection{Limit Testing}
Man betrachte eine Funktion $q: A \rightarrow W$, welche jedem konkreten Szenario eine Güte $w \in W$ zuordnet. Diese ist üblicherweise von dem Verhalten des Systems unter Test (SUT) abhängig, was bedeutet, dass das SUT in einem konkreten Szenario beobachtet wird und die Beobachtung (z.B. Simulationsergebnis) als Grundlage der Güteberechnung dienen. Das Ziel von Limit Testing ist es, die Szenarien abzutesten, für die $q$ Grenzwerte zurückliefert (siehe auch linker Pfeil in Abbildung \ref{fig:scenariosandquality}).

\subsection{\enquote{Gute} Testfälle}
Ein Testfall besteht generell aus einer Eingabe, einer erwarteten Ausgabe sowie Umgebungsbedingungen. Testszenarien (hier: konkrete Szenarien) bilden die Eingabe und die Umgebungsbedingungen. Die erwartete Ausgabe kann für die hier vorliegenden, kontinuierlichen Systeme nicht durch einen einzelnen Wert dargestellt werden, mit dem auf Gleichheit verglichen wird, sondern wird mithilfe von Bereichsangaben angegeben.

\begin{figure}[htb]
	\centering
	\includegraphics[height=4cm]{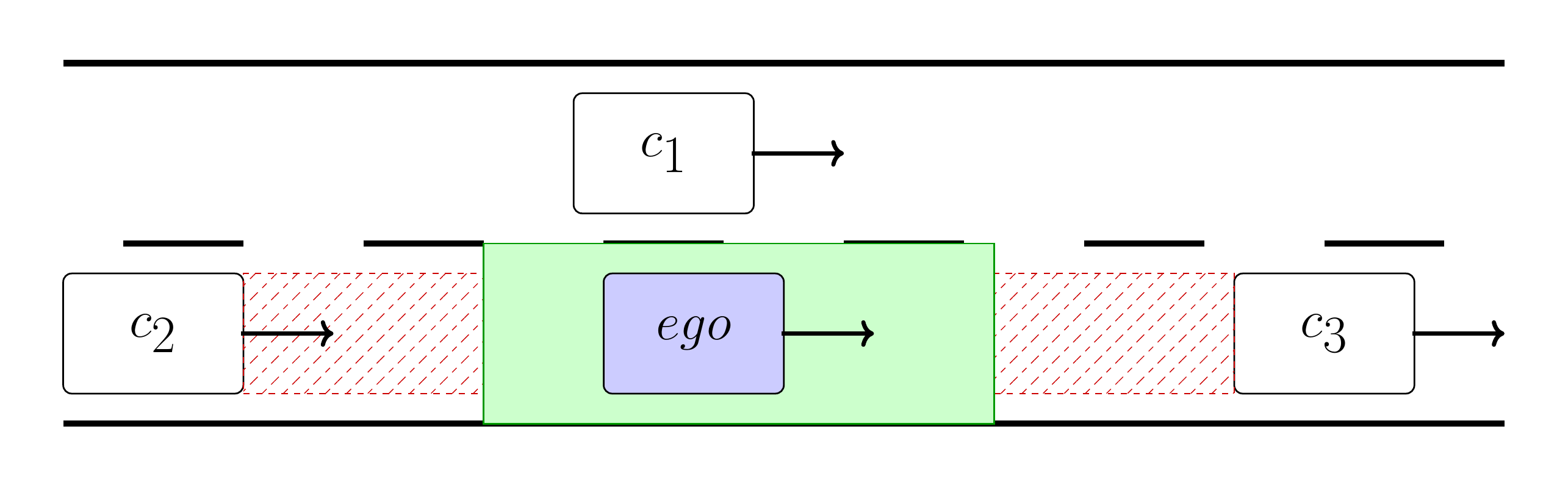}
	\caption{Beispiel für einen sicheren Operationsbereich (grün), eingeschränkt durch notwendige Abstände nach vorne und hinten (rot) sowie einem Fahrzeug auf der angrenzenden Spur ($c_1$)}
	\label{fig:safeOperationEnvelope}
\end{figure}

In dem hier betrachteten Kontext handelt es sich dabei um einen sicheren Operationsbereich (siehe Abbildung \ref{fig:safeOperationEnvelope}) \cite{koopman2016challenges}. In diesem darf das System frei handeln und solange es diesen Bereich nicht verlässt, wird es als sicher bezeichnet. Vor allem bei (möglicherweise nichtdeterministischen) Systemen, welche z.B. auf maschinellem Lernen aufbauen, spielt dies eine entscheidende Rolle, da im Vornherein nicht für jede Situation ein bestimmtes Verhalten exakt beschrieben werden kann. Im Sinne von Limit Testing wird ein \emph{guter} Testfall wie folgt definiert (siehe \cite{pretschner2015defect}):
\begin{quote}
	\centering
	\emph{Ein guter Testfall kann potentiell fehlerhaftes Systemverhalten aufzeigen.} Dies bedeutet, dass sich in einem guten Testszenario ein korrektes System den Grenzen des sicheren Operationsbereiches annähert und ein fehlerhaftes System diese überschreitet.
\end{quote}

\section{Existierende Arbeiten}\label{sec:currentapproaches}
Verschiedene Ansätze existieren in dieser Domäne. Ein sehr praktikabler Ansatz ist die Extraktion von konkreten Szenarien aus Sensordaten von Realfahrten \cite{minnerup2015collecting}\cite{zofka2015data} \cite{lages2013automatic}\cite{wolschke2017observation}\cite{zhaoandpeng} oder aus der Simulation \cite{sippl2016simulation} \cite{zofka2016testing}. Dabei stammt die Umgebung in den Szenarien aus der Realität, wird manuell erstellt oder automatisiert erzeugt \cite{kim2017testing}\cite{campos2015procedural}. Die resultierende Menge an Testszenarien kann dann anhand von Kritikalitätsheuristiken und -metriken gefiltert werden \cite{junietz2017metrik}\cite{wachenfeld2016worst}. Es kann allerdings sein, dass unter den aufgezeichneten Szenarien keine besonders \emph{guten} Testszenarien enthalten sind oder dass die enthaltenen Szenarien nicht zielgerichtet eine spezielle Funktionalität des Gesamtsystems testen. Darüber hinaus impliziert eine hohe Kritikalität nicht unbedingt, dass das jeweilige Szenario ein guter Testfall ist, da sich das System korrekt oder inkorrekt, sowohl in nicht-kritischen als auch in kritischen Szenarien, verhalten kann. Hinzu kommt, dass bei den meisten der genannten Ansätze die entstehende Menge an Szenarien nur als Open-Loop-Szenarien verwendet werden können, da eine Umwandlung in Closed-Loop-Szenarien - abhängig vom Anwendungsfall - nur sehr schwer oder überhaupt nicht möglich ist.

Ein Datenbankansatz \cite{putz2017database}\cite{putz2017system} verwendet Aufzeichnungen von Realfahrten und deren Zuordnung zu zuvor spezifizierten, logischen Szenarien \cite{roesener2016scenario}\cite{roesener2017comprehensive}. Durch eine ausreichend große Menge solcher Aufzeichnungen wird eine aussagekräftige Verteilung und Beschreibung des Realverkehrs gewonnen. Entsprechend dieser Verteilungen werden dann für den Test von neuen Systemen Parameterbelegungen gewählt, um so die häufigsten Szenarien ausreichend abzudecken. Es kommt zu den bereits genannten Problemen für Realfahrtaufzeichnungen, z.B. dass zwar viele Standardszenarien, aber wenige schwierige, das System besonders fordernde Szenarien getestet werden.

Um die Anzahl an Szenarien möglicherweise zu reduzieren, werden beobachterbasierte Ansätze vorgeschlagen \cite{otten2018automated}\cite{tatar2016test}. Diese erfordern eine spezifizierte Vorbedingung, welche den Beobachter aktiviert, und ein erwartetes Verhalten als Nachbedingung. Konkrete Szenarien werden dann so ausgewählt, dass eine hohe Abdeckung im Suchraum erzielt wird. Tatsächlich ist es jedoch überaus schwierig - wenn nicht unmöglich - diese Vorbedingungen so zu spezifizieren, dass keine Fälle vom Beobachter übersehen werden, aber auch nicht zu viele Fehlidentifikationen getätigt werden. Außerdem kann die Abdeckung eventuell nur schwer für eine große Anzahl an Szenarien erzielt werden. Eine hohe Abdeckung im Suchraum bedeutet außerdem nicht zwangsläufig, dass \enquote{gute} Testszenarien gefunden werden.  Darüber hinaus ist die Herkunft und methodische Erstellung der Bedingungen unklar.

Im Bereich der suchebasierten Verfahren in der Domäne der Fahrsysteme präsentiert die initiale Forschung die Anwendung von suchebasierten Verfahren für den funktionalen Test eines Parkassistenten \cite{buehler2003evolutionary} und eines Bremsassistenten \cite{buehler2005evolutionary}\cite{buehler2008evolutionary} (darauf aufbauend: \cite{hierlinger2017method}). In der Folge von technologischen Weiterentwicklungen wurde die Anwendung für verschiedene Steuergeräte gezeigt, basierend auf optimierten Eingangssignalen \cite{pohlheim2005evolutionary}\cite{vos2013evolutionary}\cite{hauer2017industrial}\cite{holling2016failure}\cite{matinnejad2017automated}\cite{deshmukh2017testing}. Kürzlich wurde ein weiterer technologischer Schritt veröffentlicht, welcher maschinelles Lernen verwendet, um die Szenarienoptimierung für einen Notbremsassistenten zu beschleunigen \cite{abdessalem2016testing}\\\cite{nejati2018testing}\cite{beglerovic2017testing}. Diese Arbeiten fokussieren sich auf den technologischen Aspekt der Optimierung und beschreiben lediglich für die präsentierten Fallstudien, wie dieser Ansatz angewandt werden kann. Die zugrundeliegende Methodik für die Erstellung des Suchproblems und der Zielfunktion wird nicht beschrieben. Ebenfalls bleibt der Aspekt der Vollständigkeit als auch die Einbindung eines automatischen Orakels unberücksichtigt.

\section{Suchebasierte Verfahren}\label{sec:sbapproaches}
Suchebasierte Verfahren sind mit den gegebenen Inputfaktoren -- Suchraum und Gütefunk-tion -- ein passendes Werkzeug, um das Element im Suchraum zu identifizieren, für das die Gütefunktion den (global) besten Wert zurückliefert. Im Kontext der Szenarien stellen $A \subset \mathbb{R}^n$ den Suchraum und $q$ die Gütefunktion dar. Als Ausgangspunkt dienen manuell oder zufällig ausgewählte oder bereits zuvor identifizierte, konkrete Szenarien. Nach der Simulation eines Szenarios werden die Simulationsergebnisse für die Berechnung der Güte des Szenarios bzw. des Systemverhaltens verwendet. Basierend auf den Gütewerten wählt der Optimierer anhand einer Heuristik neue konkrete Szenarien aus. Dies wird fortgesetzt, bis keine Szenarien mit besserer Güte mehr gefunden werden, eine maximale Iterationszahl erreicht oder die zur Verfügung gestellte Rechenzeit aufgebraucht ist.

\begin{figure}[htb]
	\centering
	\includegraphics[height=5cm]{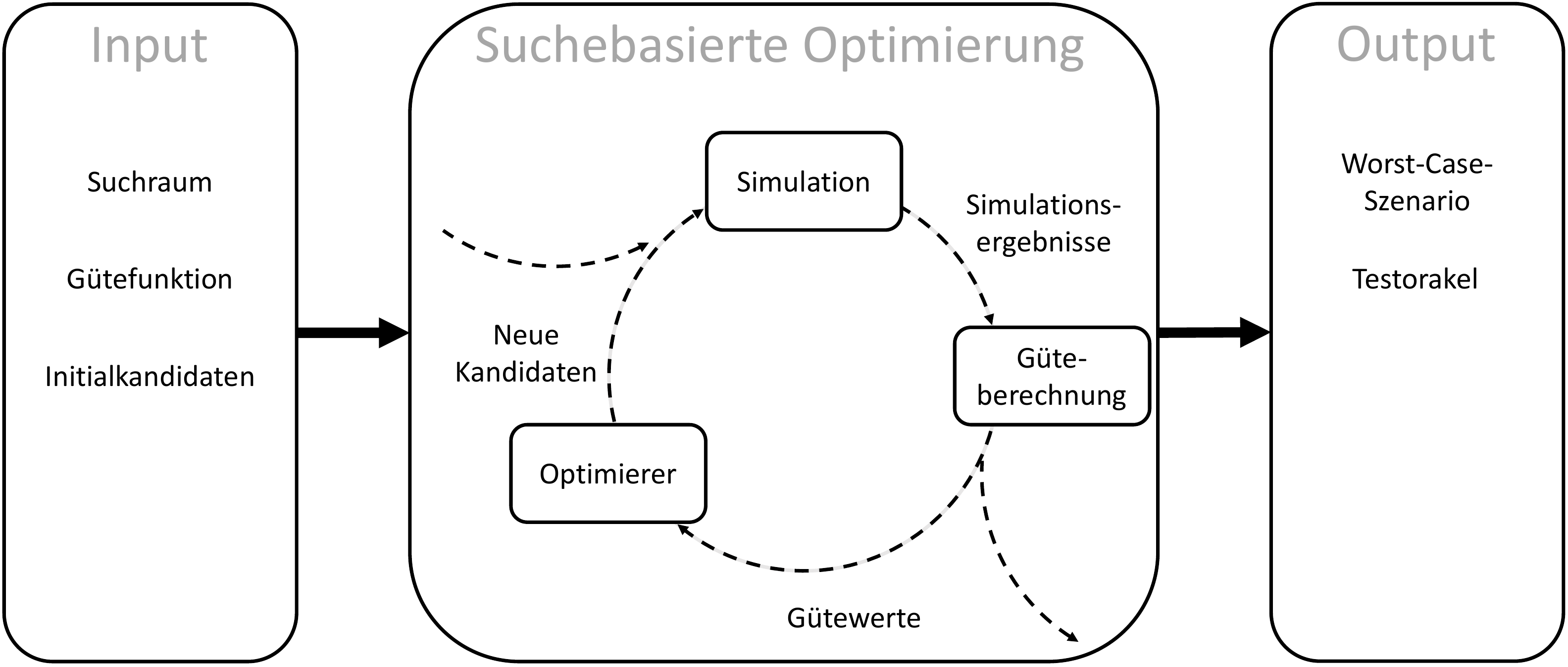}
	\caption{Anwendung von suchebasierten Verfahren auf Suchraum und Gütefunktion, um Testszenarien zu erhalten}
	\label{fig:opti}
\end{figure}

Eine Vielzahl an Verfahren und Heuristiken stehen zur Verfügung. Für diese Arbeit sind vor allem die globalen Optimierungsverfahren von Bedeutung, da - so gut wie technisch möglich - das Szenario gefunden werden soll, welches das globale Optimum darstellt. Abhängig von der Anzahl an Optimierungszielen können herkömmliche, durch die Natur und die Physik inspirierte Einziel- oder Mehrzielverfahren verwendet werden, z.B. die weitverbreiteten \cite{feldt2015broadening} genetischen Algorithmen \cite{mitchell1995genetic} (Mehrzielversion: NSGA-II \cite{deb2002fast}) und Partikelschwarmalgorithmen \cite{kennedy1995particle}. Welche Heuristik für den gegebenen Suchraum und die gegebene Gütefunktion das globale Optimum am schnellsten und zuverlässigsten findet, kann meist a-priori nicht bestimmt werden. Abhängig von den im Testprozess vorhandenen Ressourcen und der Dauer eines Simulationslaufes können unterschiedliche Arten von Verfahren gewählt werden. Für sehr begrenzte Ressourcen und besonders teure Simulationsläufe, z.B. da deutlich langsamer als in Echtzeit simuliert wird, können sequentielle Verfahren verwendet werden, die möglichst viele Informationen über den Suchraum sammeln. Diese sind nicht immer parallelisierbar. Falls ausreichend Ressourcen vorhanden sind und einzelne Simulationsläufe eher schnell ausgeführt werden können, bieten sich Verfahren an, die stark parallelisierbar sind.

\section{Methodik}\label{sec:methodology}
Existierende Arbeiten zeigen, dass suchebasierte Verfahren als Technologie in konkreten Anwendungsfällen einsetzbar sind. Die Vorstellung folgender Methodik soll nicht nur motivieren, diese Technologien als Ergänzung oder anstelle von anderen Ansätzen zu verwenden, sondern auch eine Herangehensweise bieten, um diese Technologien zielgerichtet einzusetzen, sodass eine Basis zur Argumentation über funktionale Korrektheit entsteht.

Da das vollständige Testen aller möglichen Szenarien nicht praktikabel ist, schlägt die hier vorgestellte Methodik vor, die bewährten Prinzipien der Äquivalenzklassen- und Grenzwertanalyse anzuwenden, indem zunächst die möglichen Szenarien nach Anwendungsfällen (z.B. Spurwechsel) in logische Szenarien gruppiert werden (Äquivalenzklas-sen) und anschließend für jedes logische Szenario diejenigen konkreten Szenarien identifiziert werden, für die das System mithilfe der definierten Gütefunktion am schlechtesten bewertet wird (Grenzwertbetrachtung). Diese Szenarien werden als Worst-Case-Szenarien bezeichnet. Falls das System den sicheren Operationsbereich auch in den Worst-Case-Szenarien nicht verlässt, so wird angenommen, dass dies auch in den anderen konkreten Szenarien dieses Suchraums der Fall ist. Das System wird also als sicher bezeichnet. Ein passender Suchraum und eine passende Gütefunktion sind nun so abzuleiten, dass diese Argumentation so gut wie technisch möglich unterstützt wird.

\begin{figure}[htb]
	\centering
	\includegraphics[height=7cm]{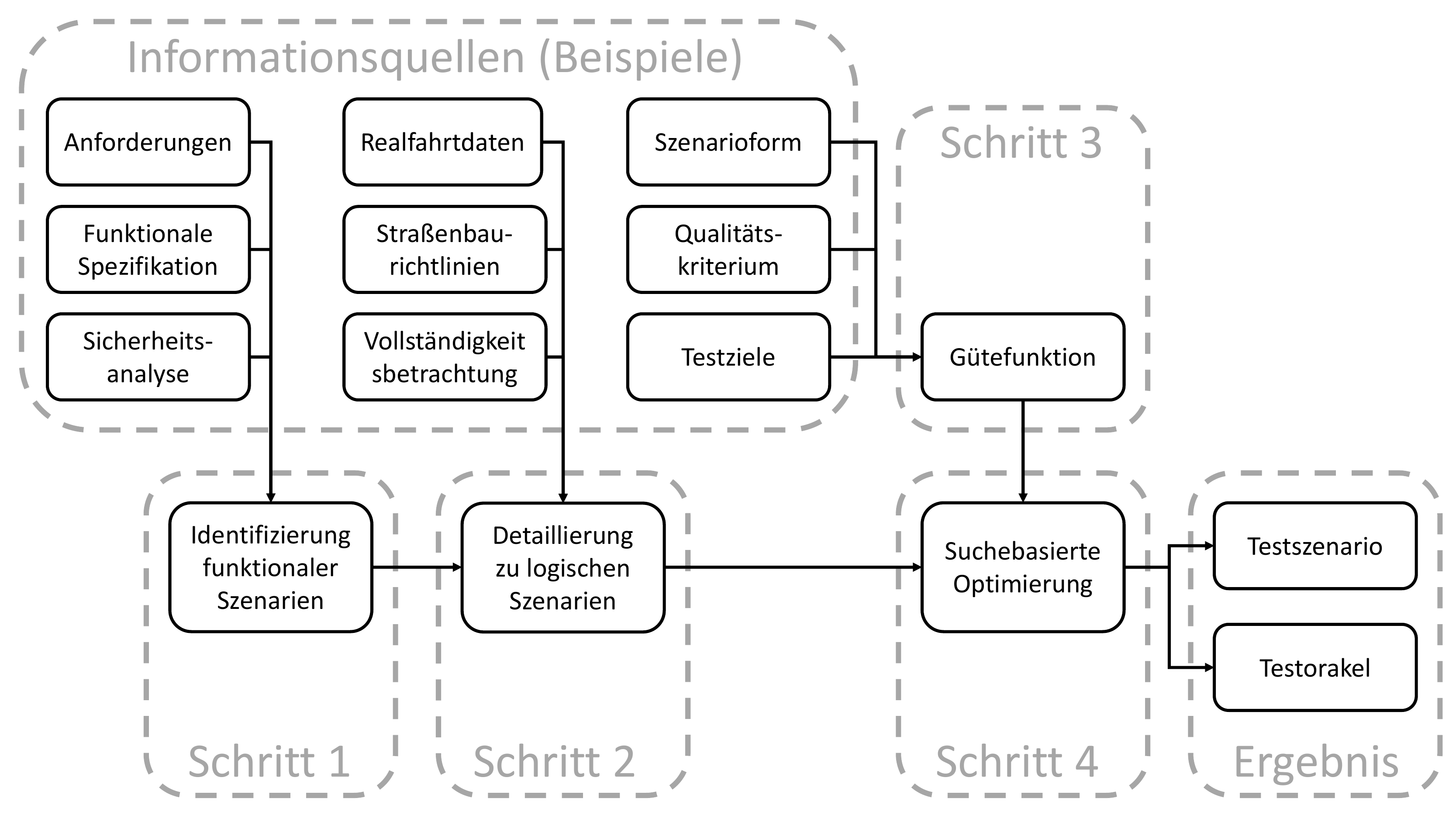}
	\caption{Überblick über die einzelnen Schritte der Methodik}
	\label{fig:methodology}
\end{figure}

Als Ausgangspunkt dienen abstrakte Informationen über das SUT, welche idealerweise in Form einer Anforderungsspezifikation oder funktionalen Spezifikation vorliegt (siehe Abbildung \ref{fig:methodology}). Zusammen mit Informationen aus Sicherheitsanalysen können funktionale Szenarien abgeleitet werden. 

Diese werden anschließend zu logischen Szenarien detailliert. Da diese den Suchraum darstellen, sind bei der Detaillierung verschiedene Aspekte zu beachten. Die fixierten Informationen zusammen mit den Variablen sollten ausreichen, damit Aussagen über das Szenario und das Systemverhalten getroffen werden können, gleichzeitig sollten diese jedoch nicht zu einschränkend gewählt werden, um keine möglichen konkreten Szenarien zu unterschlagen. Die Anzahl der Variablen sollte durch passend gewählte fixierte Informationen konstant und möglichst gering gehalten werden, um damit auch den Suchraum konstant und möglichst einfach zu halten, was die nachfolgende Suche nach dem Optimum deutlich vereinfachen kann. Für eine geeignete Wahl der Intervallgrenzen kann z.B. eine Datenanalyse von Realfahrten durchgeführt werden. Zusätzlich können (inter)nationale Richtlinien als Orientierung dienen. Beispielsweise enthält das \enquote{European Agreement on Main International Traffic Arteries} Informationen über das Aussehen von Autobahnen. Sollten sich logische Szenarien so sehr überschneiden, dass der Suchraum des einen im Suchraum des anderen (fast) vollständig enthalten ist, kann entsprechend versucht werden, die Anzahl an logischen Szenarien zu reduzieren.

Nachdem der sichere Operationsbereich durch passende Abstände oder Metriken beschrieben ist, kann eine Gütefunktion für das System aufgestellt werden. Diese kann aus mehreren Teilen zusammengesetzt werden, wobei einige Teilfunktionen mithilfe von entsprechenden Güteberechnungen sicherstellen, dass das Szenario durch die Optimierung in die gewünschte Form gebracht wird. Unter allen Szenarien, welche diese Form aufweisen, wird daraufhin das Szenario gesucht, in dem sich das System am nähesten an die Grenzen des sicheren Operationsbereichs annähert (korrektes System) oder diese Grenzen am weitesten überschreitet (fehlerhaftes System). Im aufgestellten Suchraum kann so durch Optimierung anhand der Gütefunktion nach dem Worst-Case-Szenario gesucht werden. Verlässt das System in diesem konkreten Szenario den sicheren Operationsbereich nicht, so wird angenommen, dass das System für alle konkreten Szenarien, die durch das logische Szenario dargestellt werden, sicher ist.

\section{Systemvergleich und -freigabe}\label{sec:release}
Durch methodisch korrekte Verwendung von suchebasierten Verfahren erhält man eine Liste an Gütewerten für das schlechteste Systemverhalten für unterschiedliche logischen Szenarien. Neben der Fehlerdetektion erlaubt dies auch den Vergleich von Systemen, da die Güte von bestimmten Systemfunktionen messbar gemacht wird. Beim Regressionstest bietet dies die Grundlage für eine Schlussfolgerung, ob das System ausschließlich durch Fehlerbehebung besser geworden ist oder ob an anderer Stelle ein Problem verursacht wurde. Das ist vor allem interessant, wenn erwünschte Anforderungen zu konkurrierenden Systemeigenschaften führen und ein guter Kompromiss erzielt werden muss.

Wie oben beschrieben sollen Systeme nicht den sicheren Operationsbereich verlassen, welcher durch Distanzen in Raum und Zeit beschrieben werden kann. Da eine Worst-Case-Betrachtung durchgeführt wird, wäre das beste System eines, welches immer den größtmöglichen Abstand einhält. Vor allem in dichtem Verkehr würde dies zu einem extrem zurückhaltendem Verhalten führen, was im schlimmsten Fall dazu führt, dass z.B. ein Spurwechsel unmöglich wird, da keine Lücke zwischen den anderen Autos groß genug ist. Es muss also auch überprüft werden, ob das System den vorhandenen Spielraum sinnvoll nutzt, wofür ebenfalls die vorgestellte Methodik verwendet werden kann. Es wird z.B. nach der größten Lücke auf der benachbarten Spur gesucht, bei der sich das System gegen einen Spurwechsel entscheidet.

Neben Systemversionen können aber auch verschiedene Systemvarianten verglichen werden, bei denen für die gleiche Funktionalität unterschiedliche Lösungsansätze eingesetzt werden. Für die Freigabe eines Systems können gewünschte Ziele für die einzelnen Funktionalitäten gesetzt und über die Güte überprüft werden. Da das Worst-Case-Szenario abhängig vom System ist, wird in dieser Arbeit die Verwendung einer Liste an logischen Szenarien mit den jeweiligen Gütefunktionen anstatt einer klassischen Liste von konkreten Szenarien vorgeschlagen. Tatsächlich kann es sein, dass ein konkretes Szenario, welches bei einem System ein Fehlverhalten identifiziert hat, für ein geringfügig anderes System völlig unbrauchbar ist.

\section{Zusammenfassung}\label{sec:conclusion}
Zu Beginn wurde beschrieben, dass die Adaption von suchebasierten Verfahren für das Testen von automatisierten und autonomen Fahrsystemen schwierig ist. Diese Techniken sind inhärent abhängig von einer passenden Gütefunktion und einem passenden Suchraum, deren beider Herkunft unklar ist. Zusätzlich soll eine Argumentation über die funktionale Korrektheit und über die Vollständigkeit des Testvorgangs unterstützt werden. Diese Arbeit beschreibt eine Methodik zur Verwendung solcher Verfahren, indem gezeigt wird, wie Systemverhalten bewertbar gemacht werden kann und worauf beim Erstellen des Suchproblems und beim Ableiten einer Gütefunktion zu achten ist. Es liegt in der Natur des Testens, dass keine Garantien für die Abwesenheit von Fehlern gegeben werden können, jedoch bieten die präsentierten Worst-Case-Szenarien eine Basis für eine solche Argumentation.

\cleardoublepage
\bibliographystyle{fkfs}
\bibliography{paperbib}

\end{document}